\pdfoutput=1

\documentclass[
]{ceurart}

\usepackage{listings}

\usepackage{amsthm}
\usepackage{amsmath}
\usepackage{bbm}
\usepackage[figuresright]{rotating} 
\usepackage{algorithm}
\usepackage[noend]{algpseudocode}
\usepackage{subcaption}
\usepackage{subfiles} 
\usepackage{tikz}
\usepackage{multirow}
\usepackage{colortbl}
\usepackage{makecell}

\newtheorem{definition}{Definition}

\begin{document}

\copyrightyear{2025}
\copyrightclause{Copyright for this paper by its authors.
  Use permitted under Creative Commons License Attribution 4.0
  International (CC BY 4.0).}

\conference{In: R. Campos, A. Jorge, A. Jatowt, S. Bhatia, M. Litvak (eds.): Proceedings of the Text2Story'25 Workshop, Lucca (Italy), 10-April-2025}

\title{Narrative Trails: A Method for Coherent Storyline Extraction via Maximum Capacity Path Optimization}

\author[1]{Fausto German}[%
    orcid=0009-0005-0954-4578,
    email=fgermanj@vt.edu,
    url=https://faustogerman.com,
]
\cormark[1]

\author[2]{Brian Keith}[]
\author[1]{Chris North}[]

\address[1]{Virginia Tech, Blacksburg, Virginia 24061, USA}
\address[2]{Universidad Católica del Norte, Av. Angamos 0610, Antofagasta, 1270709, Chile}

\cortext[1]{Corresponding author.}

\begin{abstract}
    Traditional information retrieval is primarily concerned with finding relevant information from large datasets without imposing a structure within the retrieved pieces of data. However, structuring information in the form of narratives---ordered sets of documents that form coherent storylines---allows us to identify, interpret, and share insights about the connections and relationships between the ideas presented in the data. Despite their significance, current approaches for algorithmically extracting storylines from data are scarce, with existing methods primarily relying on intricate word-based heuristics and auxiliary document structures. Moreover, many of these methods are difficult to scale to large datasets and general contexts, as they are designed to extract storylines for narrow tasks. In this paper, we propose Narrative Trails, an efficient, general-purpose method for extracting coherent storylines in large text corpora. Specifically, our method uses the semantic-level information embedded in the latent space of deep learning models to build a sparse coherence graph and extract narratives that maximize the minimum coherence of the storylines. By quantitatively evaluating our proposed methods on two distinct narrative extraction tasks, we show the generalizability and scalability of Narrative Trails in multiple contexts while also simplifying the extraction pipeline. The code for our algorithm, evaluations, and examples are available at \href{https://github.com/faustogerman/narrative-trails}{https://github.com/faustogerman/narrative-trails}
\end{abstract}


\maketitle

\section{Introduction}
In the last couple of decades, the fields of data science and data analytics have seen significant growth, helping people make sense of large, complex, and often interwoven data. A common task in the sensemaking process is to structure data in a format that aids analysis and information retrieval for downstream tasks \cite{pirolli2005sensemaking}. For example, structuring information in the form of narratives can help scientists communicate advanced ideas to the general public \cite{powerOfStories, kreuter_narrative_2007} and can aid with finding information in collaborative settings \cite{10.1145/2389176.2389217}. That is, narratives serve as tools for structuring complex datasets into coherent, manageable units that facilitate more effective communication and understanding of the information, ultimately reducing the cognitive load needed to make sense of information \cite{10.1145/1835804.1835884}. By organizing disparate data points into narrative structures, we enable people to identify underlying patterns, connections, and themes that might not be immediately evident by the data. For instance, placing documents in a sequential storyline may help a student researching the relationship between ``computer vision'' (CV) and ``natural language processing'' (NLP) to discover that ``image captioning'' and ``visual question answering'' bridge the concepts of CV and NLP.

However, despite the importance of narrative extraction from data, efficient algorithmic approaches are scarce, with current methods primarily relying on intricate word-based heuristics \cite{10.1145/2086737.2086744} and linear programming formulations \cite{10.1145/1835804.1835884, keith2020narrative} with limited scalability and versatility. To solve some of the scalability and availability issues of current methods, in this work we propose Narrative Trails, an algorithm for extracting coherent storylines from text documents. This helps to relate potentially disconnected ideas and extract information from large datasets. To achieve this, we approach the narrative extraction problem as a maximum capacity path optimization problem. Specifically, we utilize Dijkstra's algorithm with a MaxiMin objective to identify $k$ distinct paths that maximize the minimum coherence between documents in a sparse coherence graph representation extracted from the dataset. Following the metaphor of Narrative Maps \cite{keith2020narrative}, the Narrative Trails algorithm is analogous to route maps for hiking trails, which often provide a multiple adjacent path between a starting point and a destination.

To assess the performance of our proposed method, we present a quantitative comparison of the Narrative Trails algorithm against a random sampling, shortest simple path, and a simplified version of the Narrative Maps extraction method on two distinct tasks across four datasets. Our analyses show that the proposed approach has a lower computational cost and better performance in terms of narrative coherence.

In this paper, we make the following contributions to computational narrative extraction: (1) \textbf{Approach}: We provide a more abstractive approach to narrative extraction, focused primarily on the abstract semantic relationships between documents in a dataset; (2) \textbf{Algorithm}: We describe an efficient algorithm that merges dimensionality reduction and path optimization for coherent storyline extraction from large datasets; and (3) \textbf{Extensibility}: We provide a repository with details of our algorithm that can be used to reproduce our results or to extend our methods to other contexts beyond text.

In the next section, we review related work on narrative extraction. Section \ref{sec:methods} formalizes the Narrative Trails extraction method and its core components. Sections \ref{sec:eval} and \ref{sec:results} present and analyze our evaluation results across multiple extraction tasks. Finally, we discuss the limitations of our work and potential lines of future research in Section \ref{sec:limitations}, followed by conclusions in Section \ref{sec:conclusion}.

\section{Related Work}
\label{sec:related-work}
Computational narrative extraction lies at the intersection of artificial intelligence, natural language processing (NLP), and combinatorial optimization. The interdisciplinary nature of computational narratives makes them useful for knowledge discovery and information synthesis from large sets of data, as they allow the extraction of structured content from unstructured content.
Moreover, recent developments in NLP and deep learning models have driven computational narrative extraction to a notable area of scholarly discussion \cite{10.1007/978-3-031-28241-6_40}.

Many frameworks, models, and algorithms have been developed for computational narrative extraction, including linear sequences \cite{10.1145/1835804.1835884, 10.1145/2086737.2086744} or timelines \cite{li-li-2013-evolutionary, LIN2008473}, parallel stories \cite{xu-etal-2013-summarizing, laban-hearst-2017-newslens}, and directed acyclic graphs \cite{keith2020narrative, 10.1145/2187836.2187957, 10.1145/2339530.2339706}. Each of these revolves around the idea of storylines that weave narrative elements---sentences, documents, or clusters of documents---into coherent sequences of events. For instance, in the Narrative Maps \cite{keith2020narrative} and Metro Maps \cite{10.1145/2339530.2339706} approaches, the authors build directed acyclic graphs that interconnect multiple storylines into single narratives with potentially many starting and ending events around particular subjects. These approaches aim to extract the underlying graph structure of documents through optimization techniques that prioritize the coherence of the storylines while adhering to the global structural constraints of the narratives.

Similarly, the work of Xu et al. \cite{xu-etal-2013-summarizing} aims to build multiple timelines or storylines parallel to one another that share similarities or revolve around a complex topic. On the other hand, The newsLens algorithm \cite{laban-hearst-2017-newslens} builds multiple parallel timelines across an entire dataset that may or may not share common themes. These methods underscore the importance of narrative frameworks in providing a multifaceted view of complex topics, allowing for a richer, more nuanced understanding of events and their interconnections while acknowledging the separation in ideas between each storyline.

Finally, in the Connect the Dots approach \cite{10.1145/1835804.1835884}, the authors focus on extracting singular sequences of events that together form a narrative chain between two fixed endpoints. This method only considers one storyline at a time, displaying a linear narrative progression of the most relevant documents from a defined start to an end. They achieve this by optimizing a set of word-based linear programming constraints that maximize the coherence of the chain.

In our work, we also focus on extracting singular narrative chains. However, we focus on an abstractive approach based on the semantics of a piece of text rather than the activations of individual words within the text and their exact appearance across the narrative. This allows us to extract chains between documents that may not use the exact wording but nevertheless share similar themes. Moreover, given enough data, our abstractive approach can find smooth transitions between unrelated source and target documents.

\begin{figure}[t]
    \includegraphics[width=\textwidth]{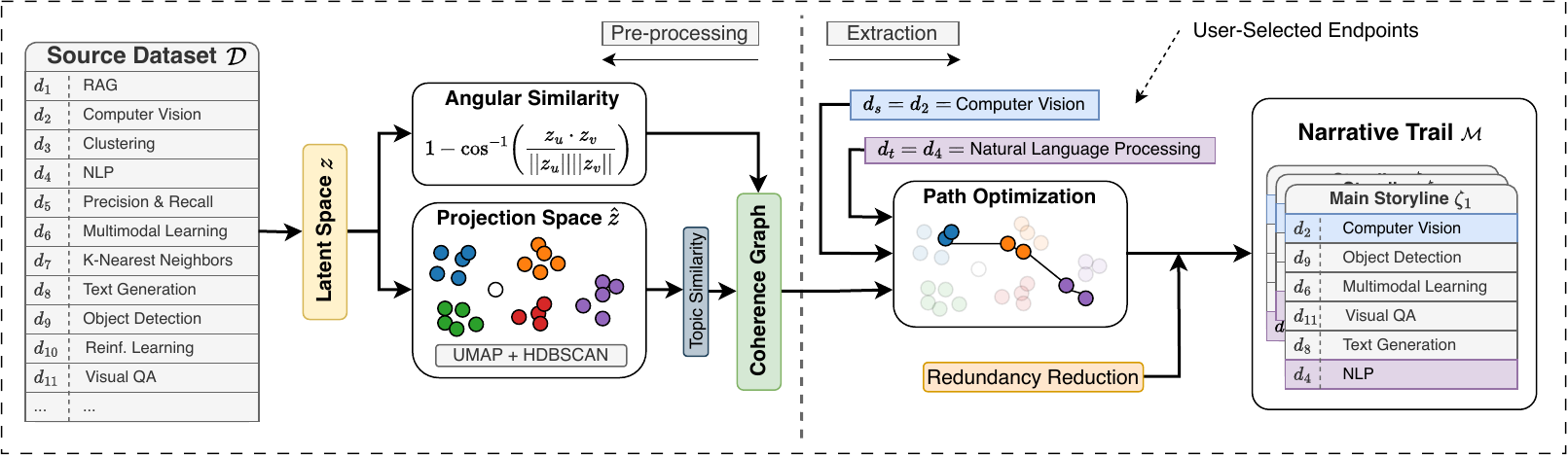}
    \caption{
        Narrative Trails extraction pipeline. Given two user-selected documents, the algorithm finds storylines that connect them with maximum capacity for coherence.
    }
    \label{fig:narrative-pipeline}
\end{figure}

\section{Methodology: Narrative Trails}
\label{sec:methods}

In this section, we introduce the formal definition of a Narrative Trail, which serves as the foundation for the structure of our narrative extraction methods.

\begin{definition}[Narrative Trail]
    \label{def:narrative-trail}
    Let $G=(V, E, w_f)$ be a weighted, directed graph built from a set of documents $\mathcal{D}=\{d_1, d_2, ..., d_n\}$, with each node $v \in V$ having an associated document $d_v \in \mathcal{D}$ and edges $(u, v) \in E$ weighted through some coherence function $w_f(u, v) = f(d_u, d_v) \geq 0$. A Narrative Trail $\mathcal{M}$ is a collection of storylines $\{\zeta_1, \zeta_2, ..., \zeta_k\}$ in $G$, where each storyline $\zeta_k=\{d_s, d_1^k, d_2^k, ..., d_l^k, d_t\}$ is an ordered sequence of documents that maximize the minimum edge weight $w_f(u, v)$ for all $(u, v) \in E$ to connect user-selected endpoints $s,t \in V$. Documents $d_1^k, d_2^k, ..., d_l^k$ are unique to their storylines.
\end{definition}

Our algorithm is a simplification of Narrative Maps \cite{keith2020narrative} for the single-pair narrative extraction case with up to $k$ distinct storylines between the source and target documents. However, unlike previous work, Narrative Trails does not rely on linear programming to extract storylines. Instead, the algorithm focuses on the latent spaces learned by deep learning models for path optimization, with the coherence graph structure and its maximum spanning tree being natural representations of the connection between documents in this space. More concretely, we divide the algorithm into projection space construction, coherence graph representation, and path optimization. Figure \ref{fig:narrative-pipeline} provides an overview of the extraction pipeline and the following subsections provide more details about each step.

\subsection{Projection Space Construction}
\label{sec:projection-space}
Our narrative extraction algorithm relies on a low-dimensional data representation constructed from the latent space learned by a deep learning model to discover topics for each document, which we achieve through a neural topic modeling method similar to BERTopic \cite{grootendorst2022bertopic}. More specifically, we use the \texttt{text-embedding-3-small} embedding model from OpenAI's API service. While OpenAI's second-generation model, \texttt{text-embedding-ada-002}, had previously shown great performance in various NLP tasks \cite{neelakantan2022text, kamalloo2023evaluating}, their latest embedding models outperform the previous models in benchmarks for English tasks and multi-language retrieval \cite{apenai_embed_2024}. We also note that other embedding models can produce similar results. For instance, the \texttt{all-mpnet-base-v2} model from the Sentence Transformers library \cite{reimers-2019-sentence-bert} has shown state-of-the-art performance in multiple embedding tasks such as semantic similarity and retrieval \cite{muennighoff-etal-2023-mteb, 10.1007/978-3-031-44204-9_1}, providing an excellent balance between speed and performance for embedding extraction. However, it is limited to a smaller context length of 384 tokens, restricting the amount of information that can be captured within the embedding representations. OpenAI's models do not have this limitation, as they allow API calls with thousands of tokens at a time.

We then use a combination of the Uniform Manifold Approximation and Projection (UMAP) algorithm \cite{mcinnes2020umap} and HDBSCAN \cite{McInnes2017} to project the data into two dimensions and assign clusters to each document. UMAP has been shown to outperform other dimensionality reduction techniques such as t-SNE \cite{tsne} and PCA by preserving more of the local and global structure of the embeddings \cite{10.1145/3428077, 10.31234/osf.io/zxvf2}. Since HDBSCAN is a density-based clustering algorithm, UMAP also serves as a complementary technique by creating low-dimensional projections where documents are more densely grouped, thereby mitigating the effects of the curse of dimensionality \cite{allaoui_considerably_2020}. Additionally, HDBSCAN's soft-clustering feature allows us to obtain cluster probability distribution vectors, which capture the probability of each document belonging to each of the discovered clusters. Comparing the topic distributions of two documents then allows us to quantify their topic similarity \cite{keith2020narrative}, which is an important component for ensuring the extracted storylines follow smooth topic transitions.

\subsection{Coherence Graph Representation}
\label{sec:coherence-graph}
Narrative Trails builds storylines by connecting documents based on the content similarity encoded in their embeddings. A naive approach to constructing the storylines is to let some storyline $\zeta_i$ be the shortest Euclidean path in the latent space between the source and target documents. However, using spatial information from the latent space alone does not provide enough structure for narrative extraction. This is because as the size of the dataset approaches infinity, finding a path between two embeddings using only their Euclidean distance would equate to finding a path of shortest distance in $\mathbb{R}^n$, which would resemble a straight line between the source and target points \cite{shortestPathRandomPoints}. However, deep learning models do not learn globally linear embedding spaces. Therefore, a straight line through the latent space may not capture the semantics of the documents well enough to define a coherent narrative. Instead, we need to define a quantity that measures how plausible it is for two documents to be connected in a storyline based on the semantics encoded in the latent space. That is, we need to define a measure for document pairwise coherence.

\subsubsection{Base Coherence}
Building on the premise that documents within a narrative should share content and context similarity \cite{10.1145/1835804.1835884, 10.1145/2487575.2487690, 8005462}, prior work \cite{keith2020narrative} defines the coherence $\theta(d_u,d_v)$ between two documents $d_u$ and $d_v$ as the geometric mean of their angular similarity in the high-dimensional embedding space and their topic similarity in the low-dimensional projection space. Formally, the coherence is defined as:
\begin{equation}
    \label{eq:base-coherence}
    \theta(d_u, d_v) = \sqrt{S(z_u, z_v)T(\hat{z}_u, \hat{z}_v)} = \sqrt{ \left( 1 - \arccos(\text{cos\_sim}(z_u, z_v)) / \pi \right) \left( 1 - \text{JSD}(\hat{z}_u, \hat{z}_v) \right)}
\end{equation}

where $z_u, z_v \in \mathbb{R}^n$ are the high-dimensional embeddings in the latent space for documents $d_u$ and $d_v$, and $\hat{z}_u, \hat{z}_v \in \mathbb{R}^m$ are their low-dimensional projections (with $m \ll n$). The angular similarity $S(z_u, z_v)$ maps the angle between the embeddings to a similarity measure in the interval $[0,1]$, and the topic similarity $T(\hat{z}_u, \hat{z}_v)$ is defined using the Jensen-Shannon divergence (JSD) between the cluster membership distributions of the documents in the low-dimensional space. We obtain the cluster membership distribution vectors from HDBSCAN during the projection space construction step. Thus, two documents are considered highly coherent with respect to one another if they exhibit \textit{both} high content similarity from $S(z_u, z_v)$ and high topic similarity from $T(\hat{z}_u, \hat{z}_v)$.

\subsubsection{Sparse Coherence}
We note that the base coherence results in a complete, undirected graph $G = (V, E, w_\theta)$, where the weights $w_\theta$ are defined by the function $\theta(d_u, d_v)$. However, we can induce sparsity into $G$ by leveraging properties of its maximum spanning tree $\text{MaxST}_G$, which is the inverse of the minimum spanning tree \cite{kleinberg2005}. Specifically, we observe that in an undirected graph, any $s$-$t$ path in the maximum spanning tree is also a path that maximizes the minimum edge weight between $s$ and $t$ in the original graph $G$ \cite{pollack_letter_1960}. This follows from the fact that $\text{MaxST}_G$ is constructed by selecting the highest-weight edges while maintaining connectivity \cite{kleinberg2005}, which ensures that for any two nodes, the weakest edge along their path is as strong as possible.

Given the equivalence between the storylines in the original graph and those in its maximum spanning tree, a tree-based approach serves as an additional optimization strategy by reducing the search space from the source node to the target during the storyline extraction phase. However, while it is possible to use $\text{MaxST}_G$ directly to identify the storylines $\zeta_i$, trees inherently provide only a single path between any two vertices. This limitation conflicts with the requirement to extract multiple storylines, as outlined in Definition \ref{def:narrative-trail}. Consequently, this structure may be too restrictive in scenarios where multiple storylines are necessary. For example, extracting the top-$k$ distinct storylines can aid in sensemaking, as each storyline may reveal a different perspective on the relationship between the endpoints. To address this, we extend the base coherence function by introducing \textit{sparse coherence}, which is defined as follows to balance sparsity with the number of possible storylines between two documents:
\begin{equation}
    \label{eq:sparse-coherence}
    \vartheta(d_u, d_v) = \mathbbm{1}\left[ u \neq v \text{ and } \theta(d_u, d_v) \geq \tau \omega \right] \theta(d_u, d_v)
\end{equation}

Where $\omega$ is the bottleneck edge weight of $\text{MaxST}_G$. The parameter $\tau \geq 0$ scales the bottleneck weight to set a hard cutoff on the minimum coherence between any two documents. It follows from Equation \ref{eq:sparse-coherence} and Kruskal's algorithm \cite{kleinberg2005} for maximum spanning trees that if $\tau > 1$, $\text{MaxST}_G$ becomes disconnected and therefore the resulting sparse graph \textit{may} also become disconnected. This is because a value of $\tau > 1$ has the effect of raising $\omega$ past the bottleneck weight that holds together $\text{MaxST}_G$. In contrast, if $\tau \leq 1$, the tree remains connected, and we can construct at least one storyline between two nodes in $G$.

\subsubsection{Incorporating Task-Specific Information}
\label{sec:task-specific-info}
While so far we have only discussed undirected graphs, Equation \ref{eq:sparse-coherence} allows us to model complex constraints through task-specific information that induce explicit directionality to the storylines. For instance, time dependencies between documents can be enforced by refining the sparse coherence definition to include an additional condition $\gamma(d_u, d_v)$ within the indicator function. The function $\gamma(d_u, d_v)$ returns \texttt{true} if and only if the date of document $d_v$ is later than that of document $d_u$, ensuring that the extracted storylines follow chronological order. To that end, the final step in the Coherence Graph Representation is to construct a weighted, directed coherence graph $G_{\vartheta} = (V, E, w_\vartheta)$, where nodes $v \in V$ represent the documents in our dataset and edges $(u, v) \in E$ with weights $w_\vartheta(u, v)$ are formed based on the sparse coherence $\vartheta(d_u, d_v)$ between documents. Specifically, an edge from node $u$ to node $v$ exists if and only if $w_\vartheta(u, v) > 0$.

\subsection{Path Optimization}
Recall that our objective is to extract a collection of $k$ distinct storylines that maximize the minimum edge weight to connect a source document $d_s$ to a target document $d_t$ in a weighted directed graph $G$. By letting $G$ equal the sparse coherence graph $G_{\vartheta}$ constructed in section \ref{sec:coherence-graph}, the extracted storylines then aim to optimize the minimum edge value or, equivalently, maximize the minimum coherence required to connect a predefined source document $d_s$ to a target document $d_t$.

The problem outlined mirrors the Maximum Capacity Path problem \cite{pollack_letter_1960}, aiming to maximize the minimum edge weight within a graph. While linear-time algorithms exist for solving this problem \cite{PUNNEN1991402}, they require additional constraints or assumptions about the graph, such as undirected graphs with distinct edges or weights in $\mathbb{N}$ \cite{KaibelPeinhardt2006}. Our coherence graph does not meet either of those requirements since it is a directed graph with real-valued edge weights from the sparse coherence between nodes. Given these constraints, we repurposed Dijkstra's algorithm \cite{dijkstra_note_1959} with a single-pair path-finding task and a maximin objective for storyline extraction. In particular, this version of Dijkstra's algorithm updates the tentative score of a node by taking the minimum between the current path's minimum edge weight and the edge weight leading to the node.

\subsubsection{Extracting \texorpdfstring{$k$}{k} Distinct Chains}
To identify the top $k$ distinct storylines from the source node $s$ to the target node $t$, we modify Dijkstra's algorithm to exclude nodes that have already been included in previously discovered paths. Specifically, after each execution of the algorithm, we record the set of nodes $V_p = \bigcup_{i=1}^{k} (\zeta_i \setminus \{d_s, d_t\})$ representing the nodes contained in each previously discovered path $\zeta_i$ and update the graph to ignore these nodes in subsequent extractions. That is, we iteratively run the modified algorithm $k$ times, each time operating on the updated graph $G'_{\vartheta}=(V \setminus V_p, E')$, where $E'$ includes only edges between the remaining nodes. By effectively ``hiding" these nodes in subsequent extractions, we prevent them from being part of any new storylines.

\subsubsection{Redundancy Reduction}
\label{sec:redundancy}
An inherent property of spanning trees is that the paths between vertices may not always be the most direct, often resulting in longer routes through the graph \cite{kleinberg2005}. This occurs because eliminating cycles reduces the number of possible shortcuts between nodes. In our pipeline, this means that the extracted storylines can sometimes be excessively long or include redundant documents. To address this issue, we implement a fast post-processing step that removes redundant documents from the storylines.

For each extracted storyline $\zeta_i$, we examine consecutive triplets of documents $(A, B, C)$ and calculate the sparse coherence values between them. We let $R$ be the geometric mean of the base coherence values $\vartheta(A, B)$ and $\vartheta(B, C)$. Since $R$ is, by definition, greater than or equal to the minimum edge weight $\omega_{\zeta}$ in the storyline, we only check whether $\vartheta(A, C) \geq R-\delta$ if the edge $(A, C)$ exists in the sparse coherence graph, where $\delta$ is a redundancy threshold parameter. If true, we consider document $B$ to be redundant and remove it from the storyline. This process creates a shortcut from $A$ to $C$ without significantly compromising the overall coherence, resulting in potentially more concise storylines.

\section{Experiments \& Evaluations}
\label{sec:eval}
In this section, we implement our proposed Narrative Trails pipeline and evaluate its performance on two distinct narrative extraction tasks to address the following questions: (RQ1) \label{rq1} How well does Narrative Trails align with human-derived shortest semantic paths? and (RQ2) \label{rq2} How do the storylines extracted by Narrative Trails compare to those extracted by the current state-of-the-art method? Our goal with these evaluations is to demonstrate the generalizability of Narrative Trails across multiple domains and tasks, as well as its adaptability to various sensemaking scenarios. Additionally, we illustrate how the flexibility of our sparse coherence definition allows us to incorporate task-specific information and constraints into the extraction pipeline.

\subsection{Evaluation Metrics}
To evaluate the intrinsic quality of the storylines, we (1) measure the minimum coherence within each storyline to verify how well our method maximizes the weakest link in the chain and (2) calculate the geometric mean of the coherence values of the edges in the storylines. We call this the ``reliability'' of the storyline as it indicates the likelihood that the documents collaboratively form a coherent storyline.

To evaluate the similarity between two storylines of potentially different lengths, we first identify the Dynamic Time Warping (DTW) path \cite{muller2007dynamic} between them using the Euclidean distance between the low-dimensional embeddings of their documents as the matching metric and normalize by the number of matches in the path (referred to hereafter as nDTW Distance), then compute the average pairwise cosine similarities between those embeddings along the resulting DTW path (referred to hereafter as DTW Similarity). Dynamic Time Warping is commonly used in time series search \cite{dtw_time_series_search} and as a metric for curve similarity \cite{dtw_curve_align}. In this context, the DTW metrics measure the semantic alignment between storylines, which we represent as curves in the low-dimensional projections. This allows us to quantify the similarity between storylines of different lengths that share related but distinct documents.

\subsection{Experimental Setup}
We used four datasets and two tasks to evaluate our proposed methods. To answer RQ1, we use a subset of human-derived paths over the Wikipedia network through the Wikispeedia game \cite{West2009WikispeediaAO}. In this game, users are tasked with finding a path between two Wikipedia pages in as few clicks as possible. Although the objective of the game---to find a shortest path---is slightly different than the goal of our proposed methods---to find a path of maximum semantic capacity---the paths extracted by humans in the Wikispeedia game have been shown to encode context about how humans perceive and explore information, especially in relation to the semantics of the network and the assigned goal page \cite{leskovecWikispeedia}. To that end, we select a subset of 10,607 finished paths from the dataset (covering 3,928 Wikipedia pages) with lengths between 7 and 20 pages per path and no back links to use as a ground truth dataset.

To answer RQ2, we use a collection of 540 news articles related to the COVID-19 pandemic and the 2021 Cuban protests \cite{keith2023iui}, 840 randomly sampled research articles from the VisPub dataset \cite{vispub}, and 1,140 randomly sampled research articles related to machine learning and AI from the AMiner dataset \cite{aminer}. These datasets feature various subtopics under a single main topic, allowing for a more focused evaluation of our experiments. In addition, we selected the AMiner and VisPub subsets at random to minimize any bias in the subtopic distributions that could provide an advantage to any of the algorithms. We implemented the Narrative Maps algorithm as a baseline by removing the coverage constraint from its linear programming formulation and setting the expected length for the main storyline extraction to 12 documents, ensuring a fair comparison with Narrative Trails.

Since OpenAI's \texttt{text-embedding-3-small} model provides 1536-dimensional embeddings, we project them to a 48 dimensions using UMAP before clustering with HDBSCAN. In all experiments, we implemented the proposed Narrative Trails algorithm and used the default values of $\tau=1$ and $\delta=0.2$ for the sparse coherence formulation and redundancy reduction, respectively. The choice of $\tau=1$ follows from Equation \ref{eq:sparse-coherence}, ensuring that the graph remains connected by the critical edge weight $\omega$. Additionally, our empirical analysis shows an average critical coherence value of $0.58\pm0.104$ across datasets. Based on this, we set $\delta=0.2$ to balance coherence and flexibility, preventing the storylines from becoming overly incoherent while allowing some variation in the documents considered redundant. In the Wikispeedia experiments, we incorporated the directed edges of the Wikipedia network as an additional task-specific constraint within the sparse coherence formulation. When comparing against Narrative Maps, we used the publication dates of the articles to enforce directionality on the edges, as detailed in section \ref{sec:task-specific-info}. Additionally, we benchmarked Narrative Trails against a random sampling method and a shortest simple path algorithm across all experimental setups.

\section{Results}
\label{sec:results}

\subsection{Alignment with Human-Derived Paths}
For our Wikispeedia evaluations, we extracted the top $k=3$ distinct paths between the source and target documents in each of the ground truth human-extracted storylines. We average the scores of the top-$k$ storylines to provide a sense of the cumulative extraction quality. Table \ref{table:wikispeedia-eval} summarizes the results of this experiment. In most cases, Narrative Trails outperforms the random sampling and simple shortest path algorithms. However, in the case of nDTW Distance, the shortest simple paths outperform our methods since it more closely models the underlying task of the Wikispeedia game by node count.

\begin{table}[t!]
    \caption{Comparison of absolute coherence and reliability, along with DTW similarity and distance for the top-$k$ extracted storylines between Narrative Trails, Narrative Trails with closeness centrality (CC), and shortest simple path using the human-derived paths from the Wikispeedia dataset as ground truth.}
    \label{table:wikispeedia-eval}
    
    \scalebox{0.81}{
    \begin{tabular}{l|ccc|ccc||ccc|ccc}
        \hline
        \multirow{2}{*}{Method} & \multicolumn{3}{c|}{Min. Coherence}              & \multicolumn{3}{c||}{Reliability}                & \multicolumn{3}{c|}{DTW Similarity}                       & \multicolumn{3}{c}{nDTW Distance}                \\
                                & $k=1$          & $k=2$          & $k=3$          & $k=1$          & $k=2$          & $k=3$          & $k=1$                          & $k=2$                          & $k=3$                                      & $k=1$          & $k=2$          & $k=3$          \\ \hline
        Wikispeedia             & 0.419          & ---            & ---            & 0.609          & ---            & ---            & ---                            & ---                            & ---                                        & ---            & ---            & ---            \\ \hline
        Random Points           & 0.320          & 0.321          & 0.322          & 0.454          & 0.455          & 0.456          & 0.347                          & 0.347                          & 0.347                                      & 2.200          & 2.201          & 2.200          \\
        Shortest Path           & 0.558          & 0.560          & 0.563          & 0.614          & 0.615          & 0.620          & 0.742                          & 0.742                          & 0.746                                      & \textbf{0.967} & \textbf{0.978} & \textbf{0.971} \\ \hline
        Narrative Trails        & \textbf{0.709} & \textbf{0.704} & \textbf{0.704} & \textbf{0.776} & \textbf{0.769} & \textbf{0.767} & \textbf{0.788}                 & \textbf{0.785}                 & \textbf{0.787}                             & 1.029          & 1.049          & 1.063          \\
        Redundancy Reduced      & 0.668          & 0.667          & 0.669          & 0.760          & 0.756          & 0.755          & 0.769\textsuperscript{$\dagger$} & 0.768                          & \underline{0.771}\textsuperscript{$\dagger$} & 1.055          & 1.076          & 1.088          \\ \hline
        Narrative Trails (CC)   & 0.640          & 0.631          & 0.630          & 0.753          & 0.748          & 0.746          & \underline{0.777}              & \underline{0.778}              & 0.766\textsuperscript{$\dagger$}             & 1.029          & 1.049          & 1.093          \\
        Redundancy Reduced (CC) & 0.630          & 0.625          & 0.624          & 0.737          & 0.735          & 0.734          & 0.759                          & 0.761\textsuperscript{$\dagger$} & 0.751                                      & 1.065          & 1.079          & 1.117          \\ \hline
    \end{tabular}
}
\end{table}

Evaluations of how humans navigate information networks using a directed $st$-task demonstrate that users typically visit hub nodes---Wikipedia pages with many incoming and outgoing links---before zeroing in on the target document \cite{leskovecWikispeedia}. Building on this observation, we investigated whether Narrative Trails could emulate similar behavior by multiplying each node's base coherence by their closeness centrality \cite{closeness_centrality} score along outgoing edges. As illustrated in Table \ref{table:wikispeedia-eval}, Narrative Trails with closeness centrality becomes the second-best performer in most cases of DTW Similarity (underlined), demonstrating the flexibility of our algorithm to approximate human sensemaking.

\subsection{Comparison with Narrative Maps}

For our comparison with Narrative Maps, we extracted the top storyline with Narrative Trails and the equivalent main storyline with Narrative Maps for 50 randomly sampled source-target pairs from each dataset. Table \ref{table:narrative-maps-eval} summarizes the results of our experiments with Narrative Trails, using Narrative Maps as a baseline algorithm. In all datasets, Narrative Trails extracts storylines with higher coherence and reliability. Moreover, when compared against the random sampling and shortest simple paths methods with the results form Narrative Maps as ground truth, Narrative Trails extracts storylines that semantically align better with the state-of-the-art Narrative Maps algorithm.

\begin{table}[t!]
\caption{Comparison of absolute coherence and reliability, along with DTW similarity and distance for the top-$k$ extracted storylines between Narrative Trails and the shortest simple path using the Narrative Maps algorithm as baseline.}
\label{table:narrative-maps-eval}

\scalebox{0.80}{
    \begin{tabular}{l|ccc|ccc||ccc|ccc}
        \hline
        \multirow{2}{*}{Method} & \multicolumn{3}{c|}{Min. Coherence}              & \multicolumn{3}{c||}{Reliability}                 & \multicolumn{3}{c|}{DTW Similarity}                        & \multicolumn{3}{c}{nDTW Distance}                                                              \\
                                & News           & VisPub         & AMnr.          & News           & VisPub         & AMnr.          & News           & VisPub                         & AMnr.          & News           & VisPub                                  & AMnr.                          \\ \hline
        Narrative Maps          & 0.499          & 0.554          & 0.502          & 0.702          & 0.677          & 0.629          & ---            & ---                            & ---            & ---            & ---                                     & ---                            \\ \hline
        Random Sample           & 0.343          & 0.412          & 0.357          & 0.492          & 0.577          & 0.512          & 0.621          & 0.337                          & 0.278          & 2.466          & 1.397                                   & 1.427                          \\
        Shortest Path           & 0.557          & 0.743          & 0.635          & 0.593          & 0.753          & 0.644          & 0.363          & 0.461                          & 0.188          & 1.001          & 0.991                                   & 1.108                          \\ \hline
        Narrative Trails        & \textbf{0.689} & \textbf{0.784} & \textbf{0.736} & \textbf{0.786} & \textbf{0.800} & \textbf{0.764} & \textbf{0.872} & \textbf{0.616}                 & \textbf{0.556} & \textbf{0.762} & \textbf{0.915}\textsuperscript{$\dagger$} & \textbf{0.962}                 \\
        Redundancy Reduced      & 0.638          & 0.756          & 0.691          & 0.739          & 0.777          & 0.724          & 0.845          & 0.570\textsuperscript{$\dagger$} & 0.455          & 0.825          & 0.946\textsuperscript{$\dagger$}          & 1.025\textsuperscript{$\dagger$} \\ \hline
    \end{tabular}
}
\end{table}

Similar to our coherence evaluation, we measured the execution time of Narrative Trails and Narrative Maps on all datasets. Since each dataset had a different number of documents, we could get a sense for how well each of the algorithms scales during the storyline extraction phase. Specifically, we measured the time it takes the algorithms to extract storylines and disregarded the time required to construct the projection space and coherence graphs. These excluded steps are considered preprocessing tasks in both algorithms, involving non-critical and cacheable operations from the end user's perspective.

\begin{figure}[t!]
    \vspace{0.25cm}
    \centering
    \includegraphics[width=0.9\textwidth]{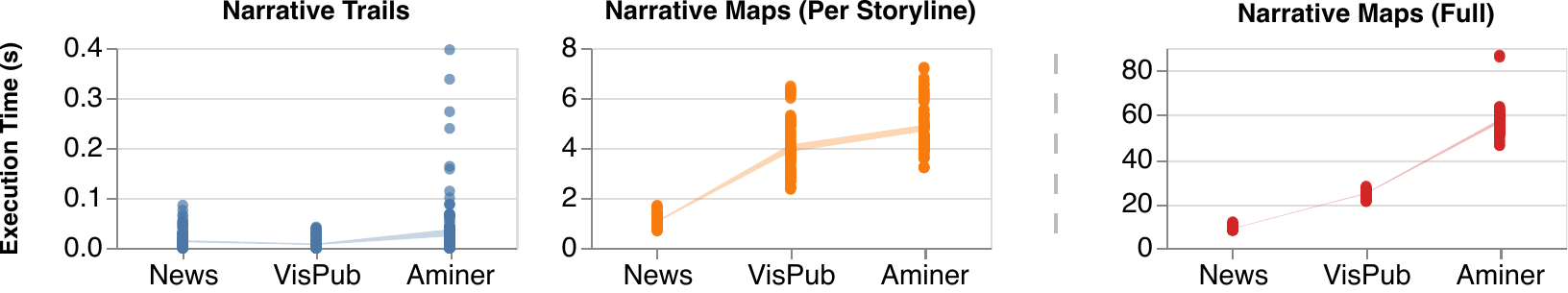}
    \caption{Comparison of extraction execution time per storyline between Narrative Trails and Narrative Maps. The diagonal lines indicate the error bands for execution time across different datasets for each algorithm.}
    \label{fig:exec-time-comparison}
\end{figure}

Our simplified narrative extraction pipeline substantially speeds up our algorithm's performance compared to Narrative Maps. Figure \ref{fig:exec-time-comparison} shows the average execution time per extracted storyline for both algorithms on the News Articles, VisPub, and AMiner datasets. We also note that since Narrative Maps is required to extract many storylines within the same narrative, we have divided the total extraction time by the number of storylines extracted (orange chart) to maintain a fair evaluation. However, we also show the execution times for the full extraction of each narrative map (red chart). These results demonstrate the efficiency of Narrative Trails when compared to Narrative Maps, especially for datasets with thousands of documents.

\begin{figure}[t!]
    \frame{\includegraphics[width=\textwidth]{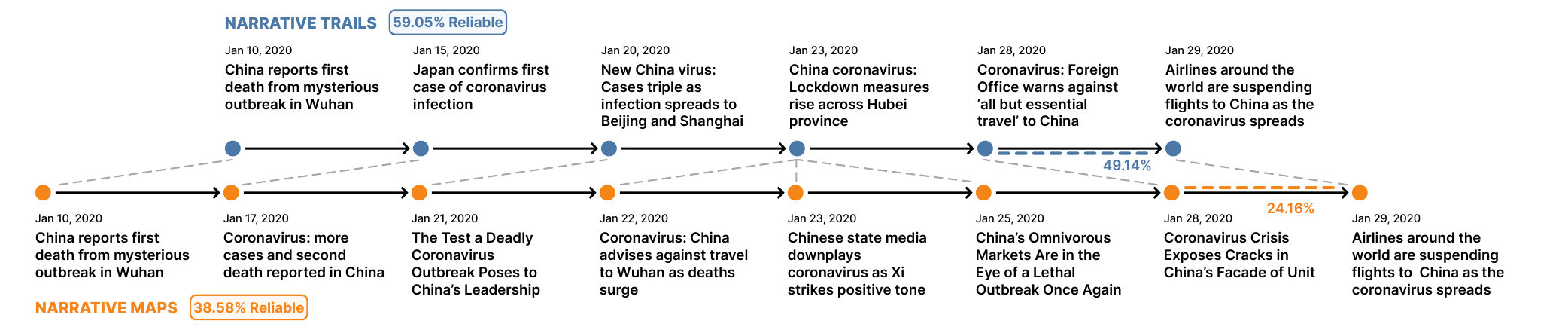}}
    \caption{Storylines about the COVID-19 pandemic's impact on global flights in January 2020, extracted from a collection of news articles using Narrative Trails (blue) and Narrative Maps (orange). The dashed gray lines represent the DTW matching between the storylines, and the dashed colored lines are the weakest links.}
    \label{fig:news-articles-case-study}
\end{figure}

Lastly, Figure \ref{fig:news-articles-case-study} showcases example storylines extracted by both algorithms to connect the first reported death from the SARS-CoV-2 virus to airlines worldwide canceling flights to China. While both storylines track the development and international response to the coronavirus outbreak, Narrative Trails emphasizes the specific developments within the health crisis that prompted an immediate response from airlines, making the storyline focused on the cause (the spread of the virus) and the effect (the suspension of flights).

\subsection{Statistical Analyses}
In most cases, our base Narrative Trails algorithm, without redundancy reduction, shows statistically significant differences across all metrics compared to other methods. However, in some instances---specifically when compared to shortest paths---the results for DTW Similarity and Distance for the redundancy reduced Narrative Trails are not statistically significant. In Tables \ref{table:wikispeedia-eval} and \ref{table:narrative-maps-eval}, we mark such cases with $\dagger$. This suggests that our redundancy reduction algorithm causes the maximum capacity path to approximate the shortest simple path between the source and target documents. Additional details on the statistical analyses are available in the linked GitHub repository.

\section{Limitations and Future Work}
\label{sec:limitations}

\paragraph{Evaluation Limitations}
We note that the evaluations with the user-extracted Wikispeedia paths as a ground-truth dataset may be difficult to interpret, as the task of the game differs slightly from the objective of narrative extraction. Thus, performing a user study or an expert-based analysis of our proposed algorithm could help to verify and contextualize its broader usability and alignment to human perception of storyline coherence.

Similarly, we do not include methods such as Connect the Dots \cite{10.1145/1835804.1835884} and newsLens \cite{laban-hearst-2017-newslens} in our evaluations. This is because the highly focused nature of these algorithms and their code availability make it difficult to incorporate them into our generalized evaluation pipeline. Additionally, our focus on single storylines in the evaluations against Narrative Maps, as opposed to the more complex narrative structures possible with other algorithms, points to future work to develop general narrative evaluation metrics that can take into account multiple interconnected storylines.

Lastly, while our evaluations report coherence---which our algorithm optimizes---as a key metric, the additional DTW metrics used in our evaluations do not rely on our optimized coherence function. Instead, DTW measures sequence similarity based on structural alignment, providing an independent assessment of storyline similarity against the Wikispeedia and Narrative Maps baselines.

\paragraph{Comparison Issues with Narrative Maps}
For a fair comparison against our methods, we removed the coverage constraint from the linear programming formulation of the Narrative Maps algorithm. The reason for this removal was twofold: (1) Narrative Trails does not include coverage constraints in its formulation, focusing solely on coherence; and (2) adding the coverage constraints is a bottleneck of the Narrative Maps extraction algorithm when the number of clusters is high, as in our evaluations, which would inflate the execution times unfairly. In practical terms, removing these constraints is equivalent to requiring a minimum average coverage of 0\% in the Narrative Maps extraction algorithm. In this context, we note that removing the coverage constraints can lead Narrative Maps to focus on a single topical cluster as it no longer needs to address the diversity requirements in topical coverage, which reduces its overall performance.

Another relevant point of comparison is that, in Narrative Maps, the extraction process can be guided by the user through semantic interactions that can directly alter its linear programming constraints. This approach provides a direct way to guide the narrative extraction and generate relevant narratives for the user. In contrast, Narrative Trails requires constraints and task-specific information to be encoded during pre-processing. Future research could explore semantic interaction models that enable users to dynamically modify and update the sparse coherence graph during extraction. For instance, deep-learning-based search agents could encode human-controllable constraints directly into its learned parameters, offering a balance between speed, accuracy, and user-defined narrative requirements as part of the semantic interaction pipeline.

\paragraph{Potential Extensions to Image Data}
Lastly, we note that the Narrative Trails algorithm is data and model-agnostic and can be extended to any context where embeddings are available. Beyond text, a trivial extension of this algorithm could extract ``concept narratives'' from image data that transition the concepts in a source image to the concepts of a target image. Applications of this extension include guided storyboarding and multi-modal narrative generation. However, these scenarios require considering the semantics of the data, as well as the representation power of the deep learning model used for embedding extraction, information retrieval, and similarity search in the context of computer vision and multi-modal learning.

\section{Conclusions}
\label{sec:conclusion}

In this paper, we presented Narrative Trails, a method for storyline extraction from large datasets with task-specific graph structures. This method addresses the limitations of current algorithms by providing an abstractive approach to narrative extraction. More specifically, we leverage the representational power of deep learning models to capture the semantics of data to form storylines. Our main insight stems from the parallels between our definition of coherent storylines and the maximum capacity path problem. The results from our experiments on human-derived paths from the Wikispeedia dataset and the comparative study with Narrative Maps demonstrate the ability of our presented algorithm to not only generalize to task-specific domains, but also to extract storylines with high coherence. Moreover, our simplified graph construction and extraction pipeline improves the overall time complexity over current methods, opening the door to future methods where extraction time is critical.

Our narrative extraction algorithm opens new avenues for future research, including contextualized global coherence and semantic interaction. One key avenue lies in improving our definition of coherence with global metrics that better guide the extraction process and regulate storyline length. Additionally, integrating deep-learning-based search agents with human-controllable constraints could improve the balance between speed, accuracy, and user-defined requirements. These improvements could further generalize our approach, broadening its applicability to context-specific domains and tasks.

Despite its limitations, our approach marks a significant contribution to computational narrative extraction, offering a more general method that simplifies the storyline extraction pipeline. Additionally, the data and model-agnostic nature of Narrative Trails opens the doors to extracting narratives from diverse datasets, including image data, thus expanding the method's applicability and impact.

\begin{acknowledgments}
Brian Keith is supported by Project 202311010033-VRIDT-UCN.
\end{acknowledgments}

\bibliography{sample-ceur}

\begin{thebibliography}{44}
\expandafter\ifx\csname natexlab\endcsname\relax\def\natexlab#1{#1}\fi
\providecommand{\url}[1]{\texttt{#1}}
\providecommand{\href}[2]{#2}
\providecommand{\path}[1]{#1}
\providecommand{\DOIprefix}{doi:}
\providecommand{\ArXivprefix}{arXiv:}
\providecommand{\URLprefix}{URL: }
\providecommand{\Pubmedprefix}{pmid:}
\providecommand{\doi}[1]{\href{http://dx.doi.org/#1}{\path{#1}}}
\providecommand{\Pubmed}[1]{\href{pmid:#1}{\path{#1}}}
\providecommand{\bibinfo}[2]{#2}
\ifx\xfnm\relax \def\xfnm[#1]{\unskip,\space#1}\fi
\bibitem[{Pirolli and Card(2005)}]{pirolli2005sensemaking}
\bibinfo{author}{P.~Pirolli}, \bibinfo{author}{S.~Card},
\newblock \bibinfo{title}{The sensemaking process and leverage points for analyst technology as identified through cognitive task analysis},
\newblock in: \bibinfo{booktitle}{Proceedings of International Conference on Intelligence Analysis}, volume~\bibinfo{volume}{5}, \bibinfo{address}{McLean, VA, USA}, \bibinfo{year}{2005}, pp. \bibinfo{pages}{2--4}.
\bibitem[{Yang and Hobbs(2020)}]{powerOfStories}
\bibinfo{author}{Y.~Yang}, \bibinfo{author}{J.~E. Hobbs},
\newblock \bibinfo{title}{The power of stories: Narratives and information framing effects in science communication},
\newblock \bibinfo{journal}{American Journal of Agricultural Economics} \bibinfo{volume}{102} (\bibinfo{year}{2020}) \bibinfo{pages}{1271--1296}. \DOIprefix\doi{https://doi.org/10.1002/ajae.12078}.
\bibitem[{Kreuter et~al.(2007)Kreuter, Green, Cappella, Slater, Wise, Storey, Clark, O'Keefe, Erwin, Holmes, Hinyard, Houston, and Woolley}]{kreuter_narrative_2007}
\bibinfo{author}{M.~W. Kreuter}, \bibinfo{author}{M.~C. Green}, \bibinfo{author}{J.~N. Cappella}, \bibinfo{author}{M.~D. Slater}, \bibinfo{author}{M.~E. Wise}, \bibinfo{author}{D.~Storey}, \bibinfo{author}{E.~M. Clark}, \bibinfo{author}{D.~J. O'Keefe}, \bibinfo{author}{D.~O. Erwin}, \bibinfo{author}{K.~Holmes}, \bibinfo{author}{L.~J. Hinyard}, \bibinfo{author}{T.~Houston}, \bibinfo{author}{S.~Woolley},
\newblock \bibinfo{title}{Narrative communication in cancer prevention and control: a framework to guide research and application},
\newblock \bibinfo{journal}{Annals of Behavioral Medicine: A Publication of the Society of Behavioral Medicine} \bibinfo{volume}{33} (\bibinfo{year}{2007}) \bibinfo{pages}{221--235}. \DOIprefix\doi{10.1007/BF02879904}.
\bibitem[{Karunakaran and Reddy(2012)}]{10.1145/2389176.2389217}
\bibinfo{author}{A.~Karunakaran}, \bibinfo{author}{M.~Reddy},
\newblock \bibinfo{title}{The role of narratives in collaborative information seeking},
\newblock in: \bibinfo{booktitle}{Proceedings of the 2012 ACM International Conference on Supporting Group Work}, GROUP '12, \bibinfo{publisher}{Association for Computing Machinery}, \bibinfo{address}{New York, NY, USA}, \bibinfo{year}{2012}, p. \bibinfo{pages}{273–276}. \DOIprefix\doi{10.1145/2389176.2389217}.
\bibitem[{Shahaf and Guestrin(2010)}]{10.1145/1835804.1835884}
\bibinfo{author}{D.~Shahaf}, \bibinfo{author}{C.~Guestrin},
\newblock \bibinfo{title}{Connecting the dots between news articles},
\newblock in: \bibinfo{booktitle}{Proceedings of the 16th ACM SIGKDD International Conference on Knowledge Discovery and Data Mining}, KDD '10, \bibinfo{publisher}{Association for Computing Machinery}, \bibinfo{address}{New York, NY, USA}, \bibinfo{year}{2010}, p. \bibinfo{pages}{623–632}. \DOIprefix\doi{10.1145/1835804.1835884}.
\bibitem[{Shahaf and Guestrin(2012)}]{10.1145/2086737.2086744}
\bibinfo{author}{D.~Shahaf}, \bibinfo{author}{C.~Guestrin},
\newblock \bibinfo{title}{Connecting two (or less) dots: Discovering structure in news articles},
\newblock \bibinfo{journal}{ACM Trans. Knowl. Discov. Data} \bibinfo{volume}{5} (\bibinfo{year}{2012}). \DOIprefix\doi{10.1145/2086737.2086744}.
\bibitem[{Keith~Norambuena and Mitra(2021)}]{keith2020narrative}
\bibinfo{author}{B.~F. Keith~Norambuena}, \bibinfo{author}{T.~Mitra},
\newblock \bibinfo{title}{Narrative maps: An algorithmic approach to represent and extract information narratives},
\newblock \bibinfo{journal}{Proc. ACM Hum.-Comput. Interact.} \bibinfo{volume}{4} (\bibinfo{year}{2021}). \DOIprefix\doi{10.1145/3432927}.
\bibitem[{Campos et~al.(2023)Campos, Jorge, Jatowt, Bhatia, and Litvak}]{10.1007/978-3-031-28241-6_40}
\bibinfo{author}{R.~Campos}, \bibinfo{author}{A.~Jorge}, \bibinfo{author}{A.~Jatowt}, \bibinfo{author}{S.~Bhatia}, \bibinfo{author}{M.~Litvak},
\newblock \bibinfo{title}{The 6th international workshop on narrative extraction from texts: Text2story 2023},
\newblock in: \bibinfo{editor}{J.~Kamps}, \bibinfo{editor}{L.~Goeuriot}, \bibinfo{editor}{F.~Crestani}, \bibinfo{editor}{M.~Maistro}, \bibinfo{editor}{H.~Joho}, \bibinfo{editor}{B.~Davis}, \bibinfo{editor}{C.~Gurrin}, \bibinfo{editor}{U.~Kruschwitz}, \bibinfo{editor}{A.~Caputo} (Eds.), \bibinfo{booktitle}{Advances in Information Retrieval}, \bibinfo{publisher}{Springer Nature Switzerland}, \bibinfo{address}{Cham}, \bibinfo{year}{2023}, pp. \bibinfo{pages}{377--383}.
\bibitem[{Li and Li(2013)}]{li-li-2013-evolutionary}
\bibinfo{author}{J.~Li}, \bibinfo{author}{S.~Li},
\newblock \bibinfo{title}{Evolutionary hierarchical {D}irichlet process for timeline summarization},
\newblock in: \bibinfo{editor}{H.~Schuetze}, \bibinfo{editor}{P.~Fung}, \bibinfo{editor}{M.~Poesio} (Eds.), \bibinfo{booktitle}{Proceedings of the 51st Annual Meeting of the Association for Computational Linguistics (Volume 2: Short Papers)}, \bibinfo{publisher}{Association for Computational Linguistics}, \bibinfo{address}{Sofia, Bulgaria}, \bibinfo{year}{2013}, pp. \bibinfo{pages}{556--560}.
\bibitem[{ren Lin and Liang(2008)}]{LIN2008473}
\bibinfo{author}{F.~ren Lin}, \bibinfo{author}{C.-H. Liang},
\newblock \bibinfo{title}{Storyline-based summarization for news topic retrospection},
\newblock \bibinfo{journal}{Decision Support Systems} \bibinfo{volume}{45} (\bibinfo{year}{2008}) \bibinfo{pages}{473--490}. \DOIprefix\doi{https://doi.org/10.1016/j.dss.2007.06.009}, \bibinfo{note}{special Issue Clusters}.
\bibitem[{Xu et~al.(2013)Xu, Wang, and Zhang}]{xu-etal-2013-summarizing}
\bibinfo{author}{S.~Xu}, \bibinfo{author}{S.~Wang}, \bibinfo{author}{Y.~Zhang},
\newblock \bibinfo{title}{Summarizing complex events: a cross-modal solution of storylines extraction and reconstruction},
\newblock in: \bibinfo{editor}{D.~Yarowsky}, \bibinfo{editor}{T.~Baldwin}, \bibinfo{editor}{A.~Korhonen}, \bibinfo{editor}{K.~Livescu}, \bibinfo{editor}{S.~Bethard} (Eds.), \bibinfo{booktitle}{Proceedings of the 2013 Conference on Empirical Methods in Natural Language Processing}, \bibinfo{publisher}{Association for Computational Linguistics}, \bibinfo{address}{Seattle, Washington, USA}, \bibinfo{year}{2013}, pp. \bibinfo{pages}{1281--1291}.
\bibitem[{Laban and Hearst(2017)}]{laban-hearst-2017-newslens}
\bibinfo{author}{P.~Laban}, \bibinfo{author}{M.~Hearst},
\newblock \bibinfo{title}{news{L}ens: building and visualizing long-ranging news stories},
\newblock in: \bibinfo{editor}{T.~Caselli}, \bibinfo{editor}{B.~Miller}, \bibinfo{editor}{M.~van Erp}, \bibinfo{editor}{P.~Vossen}, \bibinfo{editor}{M.~Palmer}, \bibinfo{editor}{E.~Hovy}, \bibinfo{editor}{T.~Mitamura}, \bibinfo{editor}{D.~Caswell} (Eds.), \bibinfo{booktitle}{Proceedings of the Events and Stories in the News Workshop}, \bibinfo{publisher}{Association for Computational Linguistics}, \bibinfo{address}{Vancouver, Canada}, \bibinfo{year}{2017}, pp. \bibinfo{pages}{1--9}. \DOIprefix\doi{10.18653/v1/W17-2701}.
\bibitem[{Shahaf et~al.(2012{\natexlab{a}})Shahaf, Guestrin, and Horvitz}]{10.1145/2187836.2187957}
\bibinfo{author}{D.~Shahaf}, \bibinfo{author}{C.~Guestrin}, \bibinfo{author}{E.~Horvitz},
\newblock \bibinfo{title}{Trains of thought: generating information maps},
\newblock in: \bibinfo{booktitle}{Proceedings of the 21st International Conference on World Wide Web}, WWW '12, \bibinfo{publisher}{Association for Computing Machinery}, \bibinfo{address}{New York, NY, USA}, \bibinfo{year}{2012}{\natexlab{a}}, p. \bibinfo{pages}{899–908}. \DOIprefix\doi{10.1145/2187836.2187957}.
\bibitem[{Shahaf et~al.(2012{\natexlab{b}})Shahaf, Guestrin, and Horvitz}]{10.1145/2339530.2339706}
\bibinfo{author}{D.~Shahaf}, \bibinfo{author}{C.~Guestrin}, \bibinfo{author}{E.~Horvitz},
\newblock \bibinfo{title}{Metro maps of science},
\newblock in: \bibinfo{booktitle}{Proceedings of the 18th ACM SIGKDD International Conference on Knowledge Discovery and Data Mining}, KDD '12, \bibinfo{publisher}{Association for Computing Machinery}, \bibinfo{address}{New York, NY, USA}, \bibinfo{year}{2012}{\natexlab{b}}, p. \bibinfo{pages}{1122–1130}. \DOIprefix\doi{10.1145/2339530.2339706}.
\bibitem[{Grootendorst(2022)}]{grootendorst2022bertopic}
\bibinfo{author}{M.~Grootendorst}, \bibinfo{title}{Bertopic: Neural topic modeling with a class-based tf-idf procedure}, \bibinfo{year}{2022}. \href{http://arxiv.org/abs/2203.05794}{{\tt arXiv:2203.05794}}.
\bibitem[{Neelakantan et~al.(2022)Neelakantan, Xu, Puri, Radford, Han, Tworek, Yuan, Tezak, Kim, Hallacy, Heidecke, Shyam, Power, Nekoul, Sastry, Krueger, Schnurr, Such, Hsu, Thompson, Khan, Sherbakov, Jang, Welinder, and Weng}]{neelakantan2022text}
\bibinfo{author}{A.~Neelakantan}, \bibinfo{author}{T.~Xu}, \bibinfo{author}{R.~Puri}, \bibinfo{author}{A.~Radford}, \bibinfo{author}{J.~M. Han}, \bibinfo{author}{J.~Tworek}, \bibinfo{author}{Q.~Yuan}, \bibinfo{author}{N.~Tezak}, \bibinfo{author}{J.~W. Kim}, \bibinfo{author}{C.~Hallacy}, \bibinfo{author}{J.~Heidecke}, \bibinfo{author}{P.~Shyam}, \bibinfo{author}{B.~Power}, \bibinfo{author}{T.~E. Nekoul}, \bibinfo{author}{G.~Sastry}, \bibinfo{author}{G.~Krueger}, \bibinfo{author}{D.~Schnurr}, \bibinfo{author}{F.~P. Such}, \bibinfo{author}{K.~Hsu}, \bibinfo{author}{M.~Thompson}, \bibinfo{author}{T.~Khan}, \bibinfo{author}{T.~Sherbakov}, \bibinfo{author}{J.~Jang}, \bibinfo{author}{P.~Welinder}, \bibinfo{author}{L.~Weng}, \bibinfo{title}{Text and code embeddings by contrastive pre-training}, \bibinfo{year}{2022}. \href{http://arxiv.org/abs/2201.10005}{{\tt arXiv:2201.10005}}.
\bibitem[{Kamalloo et~al.(2023)Kamalloo, Zhang, Ogundepo, Thakur, Alfonso-hermelo, Rezagholizadeh, and Lin}]{kamalloo2023evaluating}
\bibinfo{author}{E.~Kamalloo}, \bibinfo{author}{X.~Zhang}, \bibinfo{author}{O.~Ogundepo}, \bibinfo{author}{N.~Thakur}, \bibinfo{author}{D.~Alfonso-hermelo}, \bibinfo{author}{M.~Rezagholizadeh}, \bibinfo{author}{J.~Lin}, \bibinfo{title}{Evaluating embedding {API}s for information retrieval}, \bibinfo{year}{2023}. \DOIprefix\doi{10.18653/v1/2023.acl-industry.50}.
\bibitem[{OpenAI(2024)}]{apenai_embed_2024}
\bibinfo{author}{OpenAI}, \bibinfo{title}{New embedding models and {API} updates}, \bibinfo{year}{2024}. \URLprefix \url{https://openai.com/index/new-embedding-models-and-api-updates/}.
\bibitem[{Reimers and Gurevych(2019)}]{reimers-2019-sentence-bert}
\bibinfo{author}{N.~Reimers}, \bibinfo{author}{I.~Gurevych},
\newblock \bibinfo{title}{Sentence-{BERT}: Sentence embeddings using {S}iamese {BERT}-networks},
\newblock in: \bibinfo{editor}{K.~Inui}, \bibinfo{editor}{J.~Jiang}, \bibinfo{editor}{V.~Ng}, \bibinfo{editor}{X.~Wan} (Eds.), \bibinfo{booktitle}{Proceedings of the 2019 Conference on Empirical Methods in Natural Language Processing and the 9th International Joint Conference on Natural Language Processing (EMNLP-IJCNLP)}, \bibinfo{publisher}{Association for Computational Linguistics}, \bibinfo{address}{Hong Kong, China}, \bibinfo{year}{2019}, pp. \bibinfo{pages}{3982--3992}. \DOIprefix\doi{10.18653/v1/D19-1410}.
\bibitem[{Muennighoff et~al.(2023)Muennighoff, Tazi, Magne, and Reimers}]{muennighoff-etal-2023-mteb}
\bibinfo{author}{N.~Muennighoff}, \bibinfo{author}{N.~Tazi}, \bibinfo{author}{L.~Magne}, \bibinfo{author}{N.~Reimers},
\newblock \bibinfo{title}{{MTEB}: Massive text embedding benchmark},
\newblock in: \bibinfo{editor}{A.~Vlachos}, \bibinfo{editor}{I.~Augenstein} (Eds.), \bibinfo{booktitle}{Proceedings of the 17th Conference of the European Chapter of the Association for Computational Linguistics}, \bibinfo{publisher}{Association for Computational Linguistics}, \bibinfo{address}{Dubrovnik, Croatia}, \bibinfo{year}{2023}, pp. \bibinfo{pages}{2014--2037}. \DOIprefix\doi{10.18653/v1/2023.eacl-main.148}.
\bibitem[{Mistry and Minai(2023)}]{10.1007/978-3-031-44204-9_1}
\bibinfo{author}{D.~M. Mistry}, \bibinfo{author}{A.~A. Minai},
\newblock \bibinfo{title}{A comparative study of sentence embedding models for assessing semantic variation},
\newblock in: \bibinfo{booktitle}{Artificial Neural Networks and Machine Learning – ICANN 2023: 32nd International Conference on Artificial Neural Networks, Heraklion, Crete, Greece, September 26–29, 2023, Proceedings, Part X}, \bibinfo{publisher}{Springer-Verlag}, \bibinfo{address}{Berlin, Heidelberg}, \bibinfo{year}{2023}, p. \bibinfo{pages}{1–12}. \DOIprefix\doi{10.1007/978-3-031-44204-9_1}.
\bibitem[{McInnes et~al.(2020)McInnes, Healy, and Melville}]{mcinnes2020umap}
\bibinfo{author}{L.~McInnes}, \bibinfo{author}{J.~Healy}, \bibinfo{author}{J.~Melville}, \bibinfo{title}{Umap: Uniform manifold approximation and projection for dimension reduction}, \bibinfo{year}{2020}. \href{http://arxiv.org/abs/1802.03426}{{\tt arXiv:1802.03426}}.
\bibitem[{McInnes et~al.(2017)McInnes, Healy, and Astels}]{McInnes2017}
\bibinfo{author}{L.~McInnes}, \bibinfo{author}{J.~Healy}, \bibinfo{author}{S.~Astels},
\newblock \bibinfo{title}{hdbscan: Hierarchical density based clustering},
\newblock \bibinfo{journal}{Journal of Open Source Software} \bibinfo{volume}{2} (\bibinfo{year}{2017}) \bibinfo{pages}{205}. \DOIprefix\doi{10.21105/joss.00205}.
\bibitem[{van~der Maaten and Hinton(2008)}]{tsne}
\bibinfo{author}{L.~van~der Maaten}, \bibinfo{author}{G.~Hinton},
\newblock \bibinfo{title}{Viualizing data using t-sne},
\newblock \bibinfo{journal}{Journal of Machine Learning Research} \bibinfo{volume}{9} (\bibinfo{year}{2008}) \bibinfo{pages}{2579--2605}.
\bibitem[{Ghosh et~al.(2021)Ghosh, Nashaat, Miller, and Quader}]{10.1145/3428077}
\bibinfo{author}{A.~Ghosh}, \bibinfo{author}{M.~Nashaat}, \bibinfo{author}{J.~Miller}, \bibinfo{author}{S.~Quader},
\newblock \bibinfo{title}{Context-based evaluation of dimensionality reduction algorithms—experiments and statistical significance analysis},
\newblock \bibinfo{journal}{ACM Trans. Knowl. Discov. Data} \bibinfo{volume}{15} (\bibinfo{year}{2021}). \DOIprefix\doi{10.1145/3428077}.
\bibitem[{Sánchez-Rico et~al.(2023)Sánchez-Rico, Hoertel, and Alvarado}]{10.31234/osf.io/zxvf2}
\bibinfo{author}{M.~Sánchez-Rico}, \bibinfo{author}{N.~Hoertel}, \bibinfo{author}{J.~Alvarado},
\newblock \bibinfo{title}{Combination of cluster analysis with dimensionality reduction techniques for pattern recognition studies in healthcare data: Comparing pca, t-sne and umap}  (\bibinfo{year}{2023}). \DOIprefix\doi{10.31234/osf.io/zxvf2}.
\bibitem[{Allaoui et~al.(2020)Allaoui, Kherfi, and Cheriet}]{allaoui_considerably_2020}
\bibinfo{author}{M.~Allaoui}, \bibinfo{author}{M.~L. Kherfi}, \bibinfo{author}{A.~Cheriet},
\newblock \bibinfo{title}{Considerably {Improving} {Clustering} {Algorithms} {Using} {UMAP} {Dimensionality} {Reduction} {Technique}: {A} {Comparative} {Study}},
\newblock \bibinfo{journal}{Image and Signal Processing} \bibinfo{volume}{12119} (\bibinfo{year}{2020}) \bibinfo{pages}{317--325}. \DOIprefix\doi{10.1007/978-3-030-51935-3_34}.
\bibitem[{Hwang et~al.(2016)Hwang, Damelin, and III}]{shortestPathRandomPoints}
\bibinfo{author}{S.~J. Hwang}, \bibinfo{author}{S.~B. Damelin}, \bibinfo{author}{A.~O.~H. III},
\newblock \bibinfo{title}{{Shortest path through random points}},
\newblock \bibinfo{journal}{The Annals of Applied Probability} \bibinfo{volume}{26} (\bibinfo{year}{2016}) \bibinfo{pages}{2791 -- 2823}. \DOIprefix\doi{10.1214/15-AAP1162}.
\bibitem[{Shahaf et~al.(2013)Shahaf, Yang, Suen, Jacobs, Wang, and Leskovec}]{10.1145/2487575.2487690}
\bibinfo{author}{D.~Shahaf}, \bibinfo{author}{J.~Yang}, \bibinfo{author}{C.~Suen}, \bibinfo{author}{J.~Jacobs}, \bibinfo{author}{H.~Wang}, \bibinfo{author}{J.~Leskovec},
\newblock \bibinfo{title}{Information cartography: creating zoomable, large-scale maps of information},
\newblock in: \bibinfo{booktitle}{Proceedings of the 19th ACM SIGKDD International Conference on Knowledge Discovery and Data Mining}, KDD '13, \bibinfo{publisher}{Association for Computing Machinery}, \bibinfo{address}{New York, NY, USA}, \bibinfo{year}{2013}, p. \bibinfo{pages}{1097–1105}. \DOIprefix\doi{10.1145/2487575.2487690}.
\bibitem[{Zhou et~al.(2017)Zhou, Wu, and Cao}]{8005462}
\bibinfo{author}{P.~Zhou}, \bibinfo{author}{B.~Wu}, \bibinfo{author}{Z.~Cao},
\newblock \bibinfo{title}{Emmbtt: A novel event evolution model based on tfxief and tdc in tracking news streams},
\newblock in: \bibinfo{booktitle}{2017 IEEE Second International Conference on Data Science in Cyberspace (DSC)}, \bibinfo{address}{Shenzhen, China}, \bibinfo{year}{2017}, pp. \bibinfo{pages}{102--107}. \DOIprefix\doi{10.1109/DSC.2017.53}.
\bibitem[{Kleinberg and Tardos(2005)}]{kleinberg2005}
\bibinfo{author}{J.~Kleinberg}, \bibinfo{author}{E.~Tardos}, \bibinfo{title}{Algorithm Design}, \bibinfo{publisher}{Pearson}, \bibinfo{year}{2005}.
\bibitem[{Pollack(1960)}]{pollack_letter_1960}
\bibinfo{author}{M.~Pollack},
\newblock \bibinfo{title}{Letter to the {Editor}—{The} {Maximum} {Capacity} {Through} a {Network}},
\newblock \bibinfo{journal}{Operations Research} \bibinfo{volume}{8} (\bibinfo{year}{1960}) \bibinfo{pages}{733--736}. \DOIprefix\doi{10.1287/opre.8.5.733}, \bibinfo{note}{publisher: INFORMS}.
\bibitem[{Punnen(1991)}]{PUNNEN1991402}
\bibinfo{author}{A.~P. Punnen},
\newblock \bibinfo{title}{A linear time algorithm for the maximum capacity path problem},
\newblock \bibinfo{journal}{European Journal of Operational Research} \bibinfo{volume}{53} (\bibinfo{year}{1991}) \bibinfo{pages}{402--404}. \DOIprefix\doi{10.1016/0377-2217(91)90073-5}.
\bibitem[{Kaibel and Peinhardt(2006)}]{KaibelPeinhardt2006}
\bibinfo{author}{V.~Kaibel}, \bibinfo{author}{M.~Peinhardt}, \bibinfo{title}{On the Bottleneck Shortest Path Problem}, \bibinfo{type}{Technical Report} \bibinfo{number}{06-22}, ZIB, \bibinfo{address}{Takustr. 7, 14195 Berlin}, \bibinfo{year}{2006}.
\bibitem[{Dijkstra(1959)}]{dijkstra_note_1959}
\bibinfo{author}{E.~W. Dijkstra},
\newblock \bibinfo{title}{A note on two problems in connexion with graphs},
\newblock \bibinfo{journal}{Numerische Mathematik} \bibinfo{volume}{1} (\bibinfo{year}{1959}) \bibinfo{pages}{269--271}. \DOIprefix\doi{10.1007/BF01386390}.
\bibitem[{M{\"u}ller(2007)}]{muller2007dynamic}
\bibinfo{author}{M.~M{\"u}ller},
\newblock \bibinfo{title}{Dynamic time warping},
\newblock \bibinfo{journal}{Information retrieval for music and motion}  (\bibinfo{year}{2007}) \bibinfo{pages}{69--84}.
\bibitem[{Rakthanmanon et~al.(2012)Rakthanmanon, Campana, Mueen, Batista, Westover, Zhu, Zakaria, and Keogh}]{dtw_time_series_search}
\bibinfo{author}{T.~Rakthanmanon}, \bibinfo{author}{B.~Campana}, \bibinfo{author}{A.~Mueen}, \bibinfo{author}{G.~Batista}, \bibinfo{author}{B.~Westover}, \bibinfo{author}{Q.~Zhu}, \bibinfo{author}{J.~Zakaria}, \bibinfo{author}{E.~Keogh},
\newblock \bibinfo{title}{Searching and mining trillions of time series subsequences under dynamic time warping},
\newblock in: \bibinfo{booktitle}{Proceedings of the 18th ACM SIGKDD International Conference on Knowledge Discovery and Data Mining}, KDD '12, \bibinfo{publisher}{Association for Computing Machinery}, \bibinfo{address}{New York, NY, USA}, \bibinfo{year}{2012}, p. \bibinfo{pages}{262–270}. \DOIprefix\doi{10.1145/2339530.2339576}.
\bibitem[{Wang and Gasser(1997)}]{dtw_curve_align}
\bibinfo{author}{K.~Wang}, \bibinfo{author}{T.~Gasser},
\newblock \bibinfo{title}{Alignment of curves by dynamic time warping},
\newblock \bibinfo{journal}{The Annals of Statistics} \bibinfo{volume}{25} (\bibinfo{year}{1997}) \bibinfo{pages}{1251--1276}.
\bibitem[{West et~al.(2009)West, Pineau, and Precup}]{West2009WikispeediaAO}
\bibinfo{author}{R.~West}, \bibinfo{author}{J.~Pineau}, \bibinfo{author}{D.~Precup},
\newblock \bibinfo{title}{Wikispeedia: an online game for inferring semantic distances between concepts},
\newblock in: \bibinfo{booktitle}{Proceedings of the 21st International Joint Conference on Artificial Intelligence}, IJCAI'09, \bibinfo{publisher}{Morgan Kaufmann Publishers Inc.}, \bibinfo{address}{San Francisco, CA, USA}, \bibinfo{year}{2009}, p. \bibinfo{pages}{1598–1603}. \URLprefix \url{https://dl.acm.org/doi/10.5555/1661445.1661702}.
\bibitem[{West and Leskovec(2012)}]{leskovecWikispeedia}
\bibinfo{author}{R.~West}, \bibinfo{author}{J.~Leskovec},
\newblock \bibinfo{title}{Human wayfinding in information networks},
\newblock in: \bibinfo{booktitle}{Proceedings of the 21st International Conference on World Wide Web}, WWW '12, \bibinfo{publisher}{Association for Computing Machinery}, \bibinfo{address}{New York, NY, USA}, \bibinfo{year}{2012}, p. \bibinfo{pages}{619–628}. \DOIprefix\doi{10.1145/2187836.2187920}.
\bibitem[{Keith~Norambuena et~al.(2023)Keith~Norambuena, Mitra, and North}]{keith2023iui}
\bibinfo{author}{B.~F. Keith~Norambuena}, \bibinfo{author}{T.~Mitra}, \bibinfo{author}{C.~North},
\newblock \bibinfo{title}{Mixed multi-model semantic interaction for graph-based narrative visualizations},
\newblock in: \bibinfo{booktitle}{Proceedings of the 28th International Conference on Intelligent User Interfaces}, IUI '23, \bibinfo{publisher}{Association for Computing Machinery}, \bibinfo{address}{New York, NY, USA}, \bibinfo{year}{2023}, p. \bibinfo{pages}{866–888}. \DOIprefix\doi{10.1145/3581641.3584076}.
\bibitem[{Isenberg et~al.(2017)Isenberg, Heimerl, Koch, Isenberg, Xu, Stolper, Sedlmair, Chen, M{\"o}ller, and Stasko}]{vispub}
\bibinfo{author}{P.~Isenberg}, \bibinfo{author}{F.~Heimerl}, \bibinfo{author}{S.~Koch}, \bibinfo{author}{T.~Isenberg}, \bibinfo{author}{P.~Xu}, \bibinfo{author}{C.~Stolper}, \bibinfo{author}{M.~Sedlmair}, \bibinfo{author}{J.~Chen}, \bibinfo{author}{T.~M{\"o}ller}, \bibinfo{author}{J.~Stasko},
\newblock \bibinfo{title}{vispubdata.org: A metadata collection about {IEEE} visualization ({VIS}) publications},
\newblock \bibinfo{journal}{IEEE Transactions on Visualization and Computer Graphics} \bibinfo{volume}{23} (\bibinfo{year}{2017}) \bibinfo{pages}{2199--2206}. \DOIprefix\doi{10.1109/TVCG.2016.2615308}.
\bibitem[{Tang et~al.(2008)Tang, Zhang, Yao, Li, Zhang, and Su}]{aminer}
\bibinfo{author}{J.~Tang}, \bibinfo{author}{J.~Zhang}, \bibinfo{author}{L.~Yao}, \bibinfo{author}{J.~Li}, \bibinfo{author}{L.~Zhang}, \bibinfo{author}{Z.~Su},
\newblock \bibinfo{title}{Arnetminer: extraction and mining of academic social networks},
\newblock in: \bibinfo{booktitle}{Proceedings of the 14th ACM SIGKDD International Conference on Knowledge Discovery and Data Mining}, KDD '08, \bibinfo{publisher}{Association for Computing Machinery}, \bibinfo{address}{New York, NY, USA}, \bibinfo{year}{2008}, p. \bibinfo{pages}{990–998}. \DOIprefix\doi{10.1145/1401890.1402008}.
\bibitem[{Freeman(1978)}]{closeness_centrality}
\bibinfo{author}{L.~C. Freeman},
\newblock \bibinfo{title}{Centrality in social networks conceptual clarification},
\newblock \bibinfo{journal}{Social Networks} \bibinfo{volume}{1} (\bibinfo{year}{1978}) \bibinfo{pages}{215--239}. \DOIprefix\doi{https://doi.org/10.1016/0378-8733(78)90021-7}.

\end{thebibliography}

\end{document}